# How to Implement Access Rights in an MIS Project

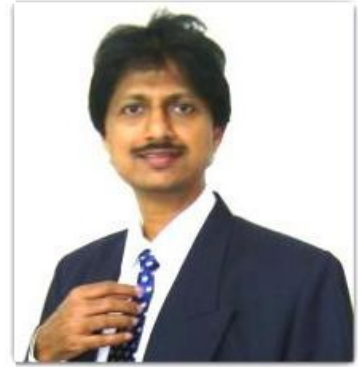

**By- Umakant Mishra, Bangalore, India**
umakant@trizsite.tk, http://umakant.trizsite.tk

## Contents

## 1. The risks involved with the Technology

An MIS project may be designed to be implemented through a LAN or through an Intranet. A LAN based MIS may work in client server models while an Internet/ Intranet based MIS may work from an web server using TCP/IP protocols. There are advantages and disadvantages of both of these implementations.

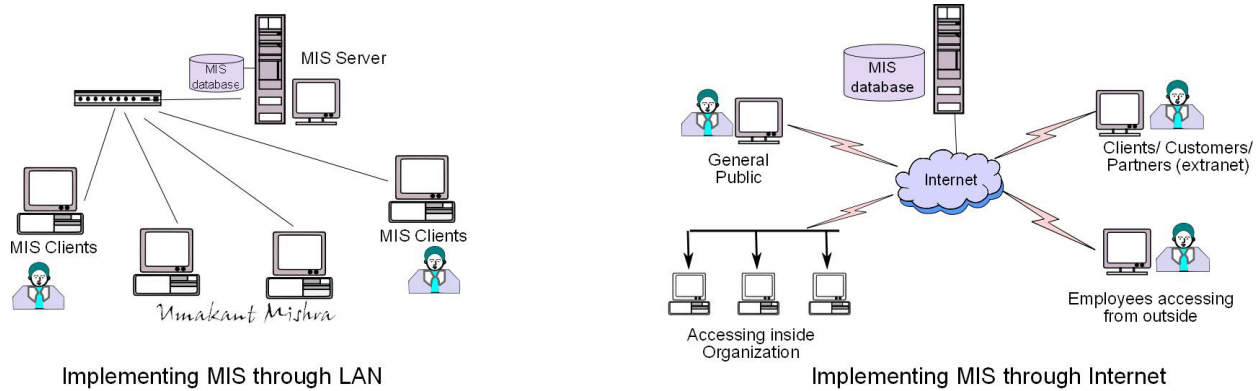

Implementing MIS through LAN        Implementing MIS through Internet

**LAN based MIS**- if you have most of the management staff located in the head office, then it will be good to implement a LAN based MIS. **Advantages**- easy to develop the MIS software, easy to manage data security, faster access from the local server and no dependency on external server or Internet connection. **Disadvantages**- cannot be accessed from outside.

**Internet based**- if you have offices at different locations, or people in senior management are mostly traveling then a LAN based MIS system is inefficient. A LAN based MIS cannot be accessed from outside the LAN. Hence, it is better to place the data on an Internet based server in order to make it accessible from outside. **Advantages**- The data can be accessed/ updated from different locations (from anywhere in the world). **Disadvantages**- difficult to program an Internet based MIS, difficult to maintain security, cannot be accessed without Internet connectivity.

**Other types**- there can be other methods of implementing MIS. One hybrid method is to maintain a LAN based MIS and update the summary data periodically on the Internet. Other implementations are broadly similar to or are derived from the above two implementations.

### Risk of keeping data on Internet

Today the Internet is like a busy street or market place where you find almost everything you want. There are all kinds of people, both good and bad, interacting with Internet. Hence, there are both advantages and disadvantage of keeping MIS data on the Internet. While Internet provides easy accessibility of data to different people from different locations, it also creates high degree of vulnerability to the valuable/ sensitive/ confidential MIS data of the organization. Unless protected properly the critical organizational data may be unduly accessed by:

- Wrong people in the organization
- Staff who have already left the organization
- Competitors and rivals in the business
- Malicious codes like spam, viruses, Trojans etc.
- Malefic people like hackers etc.

## 2. Analysis of Data Access Rights

An MIS is generally intended to be used by the higher level of management. But in some case the MIS may have some extended modules to be accessed by middle level managers or even common general staff. However, the data that is meant to be accessed by the higher level of management are generally not required to be accessed by other general staff. As some part of the MIS data may be sensitive and confidential, it is important to implement well thought access rights to ensure that the valuable organizational data is being protected from misuse by wrong hands and the right portion of the data is available to the right people.

### Categorizing users as employee and outsider

From one perspective the world of Internet users may be divided into two groups, the employees of an organization and outsiders. Prima-facie it is viewed that the organizational data should be available only to the people working inside the organization and blocked from the outsiders. However, this assumption is not always true. A further analysis explores some exceptions to this concept. For example,

- Every staff of the organization doesn't need access to the critical managerial data or sensitive organizational data.
- Some people from outside (such as key customers, partners, funding agencies etc.) may be given limited access to specific portion of the data.

## Defining hierarchy of Staff

In order to define different access levels it is often important to look at the organizational structure. The structure of an organization may be hierarchical or matrix or hybrid. In a hierarchical structure the CEO or ED may occupy the top position in the hierarchy followed by a group of directors/ GMs heading different activities of the organization. The following diagram represents a symbolic structure of an organization (typically an NGO) working in developmental sector. This organogram depicts a schematic view of the hierarchy and does not elaborate the exact designations or number of staff under different categories.

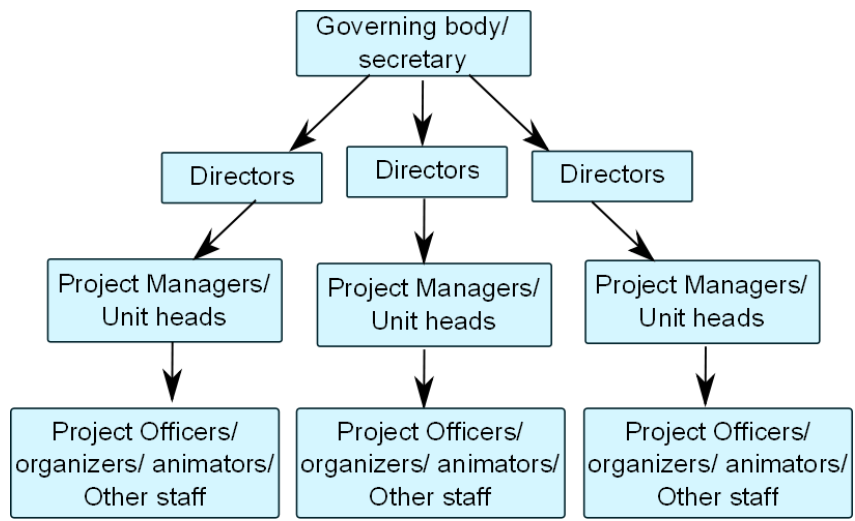

The above diagram shows roughly three layers of management, viz. secretary, directors and managers, out of which the later two are more involved in project implementation and day-to-day activities of the organization. As the MIS data is designed to support the managers in taking effective decisions, there is no need to expose that kind of data to the common general staff as because that may be unnecessarily disturbing to the employees of that level.

## Categorizing data according to criticality/ confidentiality

It is not enough just to categorize the levels of management and decide which levels of managers are eligible to access the MIS data because all the data don't bear equal significance. Hence it is important to categorize the data from different angles like its criticality, sensitivity and confidentiality.

### 1. General organizational data (non-critical, non-confidential)
⇨ Such as who are the staff in different units of the organization
⇨ What are the current projects or products

## 2. Managerial data (critical but not-confidential)
- ⇨ Monthly progress reports for each project
- ⇨ Monthly financial expenses on different heads
- ⇨ Monthly sales figures etc.

## 3. Sensitive and Confidential data
- ⇨ Performance appraisal data
- ⇨ Internal recommendations for staff training and promotions
- ⇨ New project proposals under negotiation
- ⇨ Staff complains/ grievances

While the general organizational data may be accessed by any staff of the organization, the critical managerial (non-confidential) data may be accessed by any manager, director and above. But the third category of data (sensitive and confidential) may be accessible only to specific senior members in the organization.

### Combining the above two Categories

As we saw above, it is not enough to classify people who will be accessing the MIS data; it is also necessary to classify the data based on its criticality and confidentiality in order to provide right data to the right people.

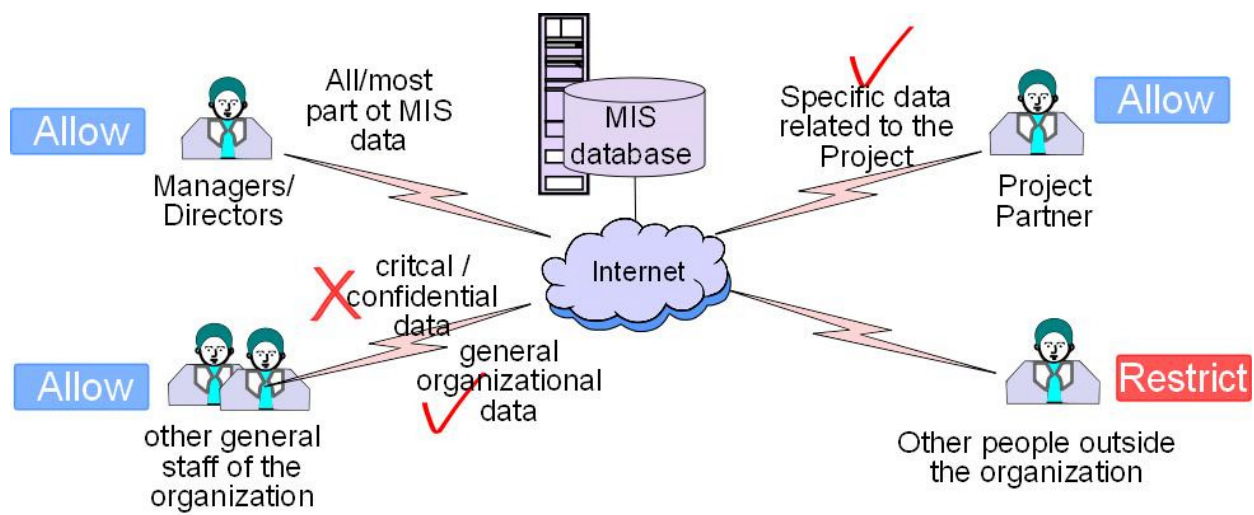

Implementing Access Rights in MIS

The above illustration gives a more elaborative picture of people who should have access to an organizations MIS data and who should not have. To summarize,

- All the employees of the organization may have access to the common organizational data such as staff positions, project related data and other organizational activities. But most employees don't need access to the critical and confidential data.
- Some outsiders like key customers or project partners may have access to specific portion of MIS data.

**Level of access from technology point of view**

Apart from the above two classifications (type of people and type of data), there is also another angle of looking at the access from technical point of view. The level of access to any data may be classified as;

- (Reading) Just viewing the data
- (Writing) Adding/ modifying the data
- (Administering) or Deciding who can do what

There are great differences between these three levels of data access. While the first level (viewing) cannot change the data, the second level (writing) can change the data, and the third level (administration) can change rights to access the data.

## Risks involved at different level of access

Allowing wrong people for reading can leak the confidential organization data, allowing wrong people for writing can mess the data with garbage and allowing wrong people to administer can erase the data or make the MIS inaccessible.

## Determining access rights by combining all the above

Although the MIS data is meant for the people in the higher level of Management we found many considerations for implementing access rights and restrictions.

- Access according to the role of the staff in the organization- some portion of the data may be accessed by managers, common general staff and some portion of data may be accessible by key customers and partners.
- Access according to the nature of data- While the non-confidential managerial data may be accessed by all levels of managers, the confidential or sensitive data may be accessible only to specific key positions.
- Access levels (whether to read, write or administer)- While a large number of MIS users will be allowed to view, only a few will be allowed to update the data. The administration rights may be confined to only one person with necessary backup arrangements for emergencies.

## 3. Implementing the Access Control

There are many methods of controlling access rights for different individual users. One method is to create an access table defining which individual user can access which individual report (or form). This method has the advantage that it enables the MIS administrator to allow specific access rights to specific individual users. But it is an elaborate method and suffers from the following drawbacks.

- ⇨ Linking every user (say about 100) to every report (say about 100 or 1000) will lead to enormous number of records on access rights.
- ⇨ Manipulating too many records for each individual not only creates too much administration load, but also leads to confusion and administrative errors.

Hence there is a need to achieve the precision of above control mechanism through an easy method. One of the methods that may suite to most organizations is to first classify the users of the organization and then define the exceptions. The roles / actors in the MIS of an organization may be defined as below.

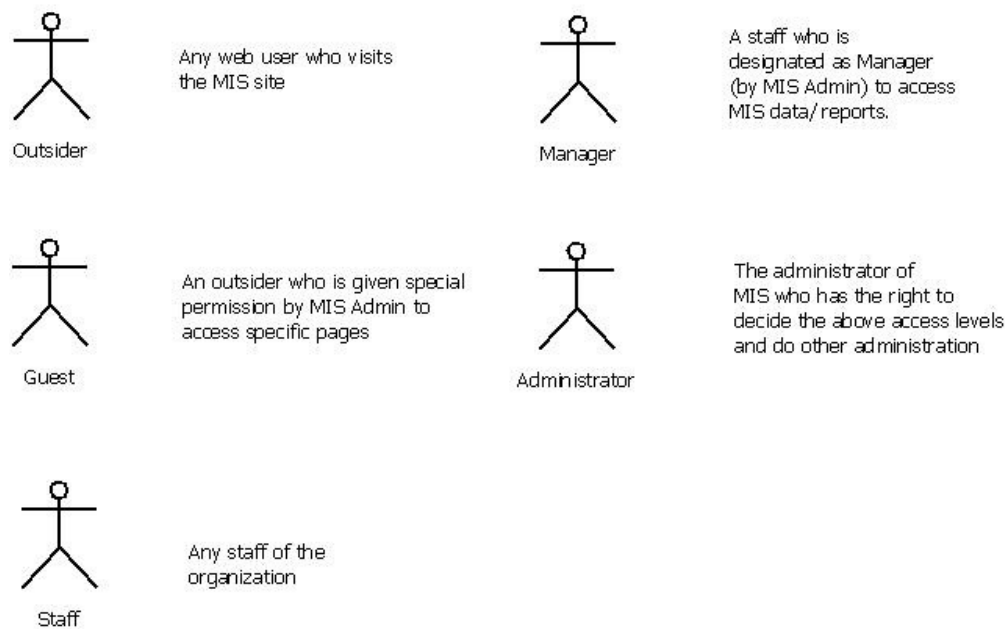

Defining actors in an MIS

The above illustration shows 5 categories of users, (1) outsider, (2) guest, (3) staff, (4) manager, (5) administrator. The "outsider" and "guest" are not employees of the organization while the "staff", "managers" and "administrators" are employees of the organization. The "outsider" will have minimum access to the MIS while "administrator" will have maximum access to the MIS.

### Defining users

An **outsider** is anybody who visits the MIS website without having any special access rights. He can view only some common pages about the organization or about the site. Those pages will contain non-critical and non-confidential general info for the public.

A **guest** is an outsider who has a valid username/ password to login. A guest may have access to specific pages. The MIS administrator will decide the access rights of the guest. However the guest should have minimal rights and not have access to data entry or administration.

A **staff** is a bonafide employee of the organization. Every staff may have access to certain organizational data available for the staff.

A **manager** is a staff who is privileged to view the MIS reports. As a manager is by default a staff, he should also be allowed to view the pages meant for the staff. The manager here is different from the designation "manager" in the organizational structure. Generally the directors, general managers and other senior managers may be designated as managers (in the MIS). The MIS admin will determine the managers based on their role in the organization.

An **administrator** is a staff who is well versed with system administration and controlling access rights. Typically the IT/IS manager is given the responsibility of MIS administration. An administrator is often called as GOD and the MIS administrator will have access to all the modules including data entry and MIS administration.

### MIS Menu use case diagram

The access to the MIS data, whether read or write or admin, will generally be routed through menus (or links or buttons or similar). As the access requirements of different types of users are different it will be useful to create different menu groups (such as, staff menu, manager menu, admin menu etc.) for different types of users (staff, manager, admin etc.). The following diagram shows the access rights of different types of users to different groups of menu.

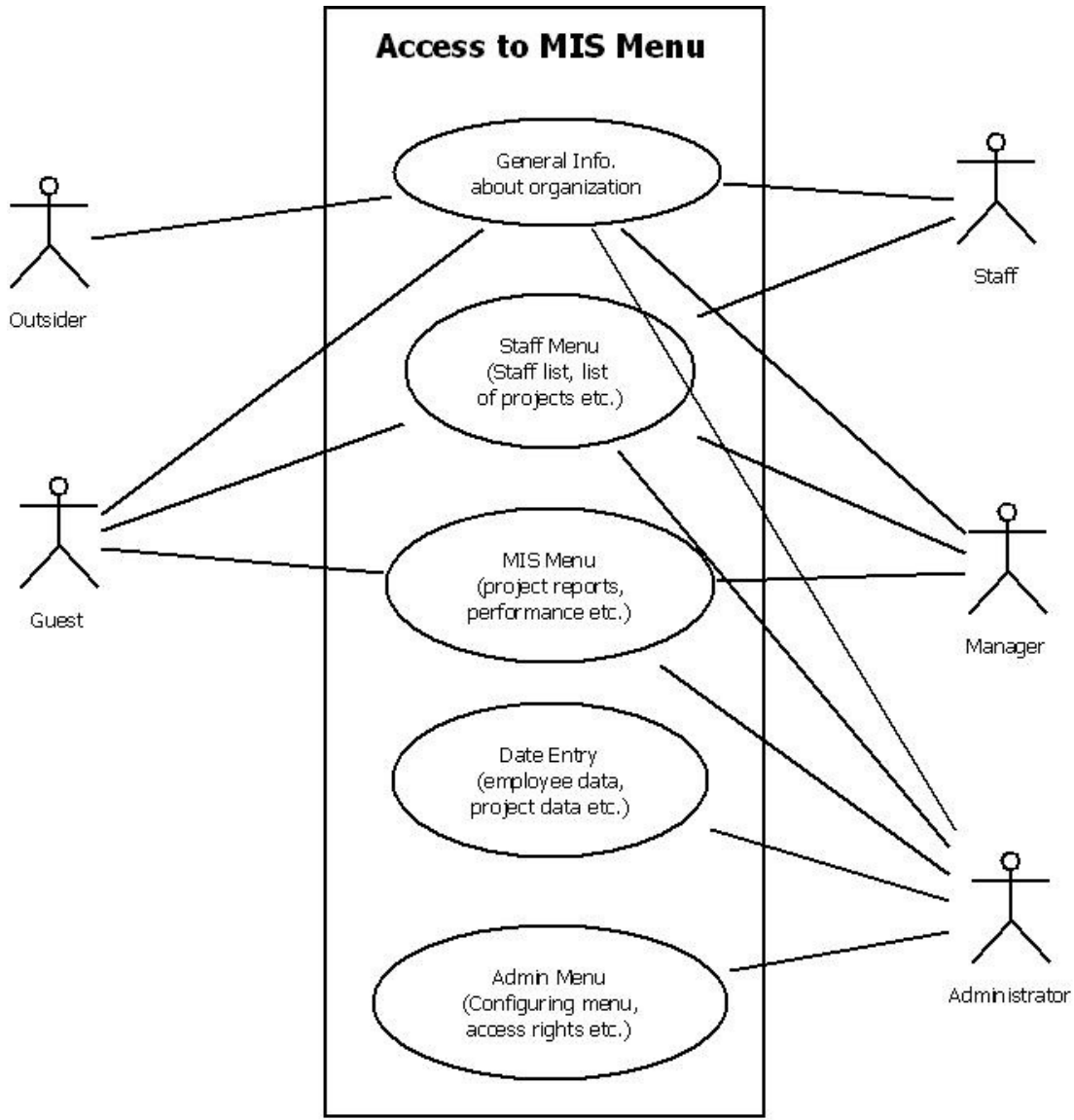

### Creating new users

Generally the MIS administrator will create the users and allocate the passwords. Then the MIS administrator will designate a role (such as, guest, staff, manager etc.) to the user. The user can access the MIS depending upon his designated role.

Many websites allow users to self-register. But this mechanism may not be suitable for most MIS sites. As the MIS site is an intranet site all the users belong to a known circle. Hence, an outsider may not[1] be allowed to register himself as a user.

---

[1] As MIS is implemented in an intranet environment there will be more disadvantages in allowing self-registration than advantages. For example, the number of self registered members may unnecessarily go up increasing burden on the MIS administrator. Besides the method needs to be protected from web robots from making spam entries.

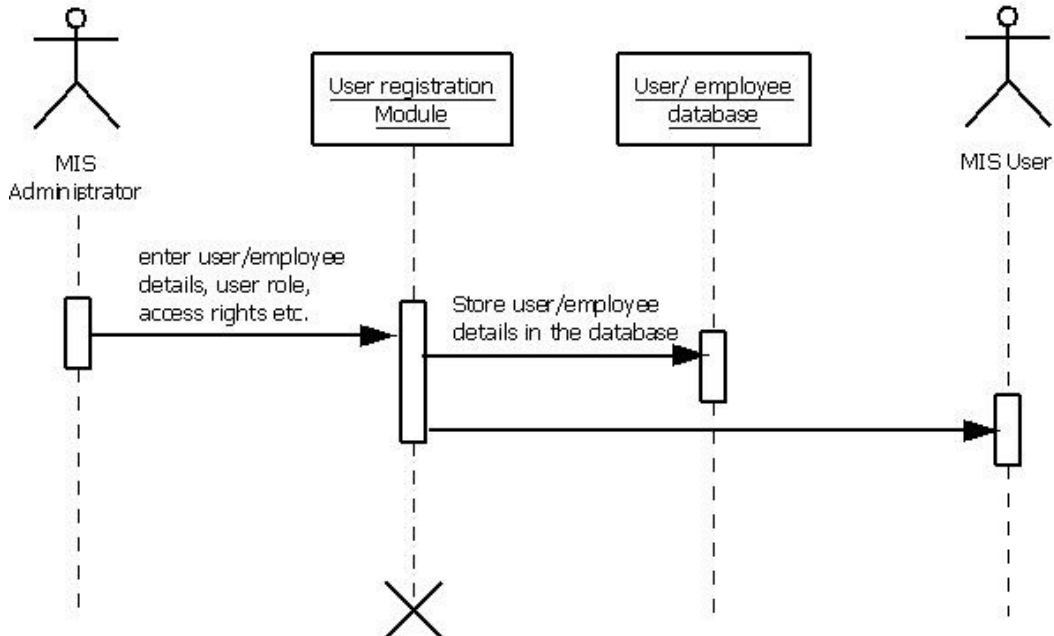

**Creating Users by MIS Administrator Sequence Diagram**

However, there may be exceptions. For example there may be a large number of customers who are supposed to view specific data for a temporary period of relevance. In such cases it is not possible for the administrators to create so many users. Hence the users may be allowed to self-register.

The risk of self registration may be controlled by automating specific access or allowing access only after the admin designates him a role or specific access to view specific restricted pages. The Administrator may allow special access to any user or may disable a user as and when required. Once registered, the users may nor may not have rights to change their passwords depending the IT policy of the organization.

### Role of MIS administrator

Role of MIS administrator will be critical to decide the following:

⇨ Which staff should be designated as manager (or admin[2])

⇨ Which outsider should be designated as guest

⇨ Which items should be placed under staff menu (for general staff)

⇨ Which items should be placed under MIS Reports (for managers)

⇨ Which user (staff/ guest/ manager etc.) will have special access to view which report or enter which data.

### The uniqueness of the User ID

As the users will be authenticated by their user-id and password, the user-id for each user should be unique. This may be ensured by putting a validation mechanism. If the administrator tries to enter a user-id that already exists the database then the MIS system should give a warning and prompt the administrator to enter a different user-id. The following logic may be used to validate user-id and ensure unique user-id for each individual MIS user.

---

[2] Having multiple administrators has its own advantages and disadvantages. The major advantage is that the backup administrator can do necessary administrator in the absence of primary administrator. However, the major disadvantage is that any of the administrators may mesh up the situation and blame each other. This policy may be decided according to the company IT policy. For example, if there are more than one administrators for server admin/ website admin etc. then there may be more than one admin for MIS admin. Otherwise it is safe to have only one admin and keep the password in an envelop with ED which can be used in emergency situations.

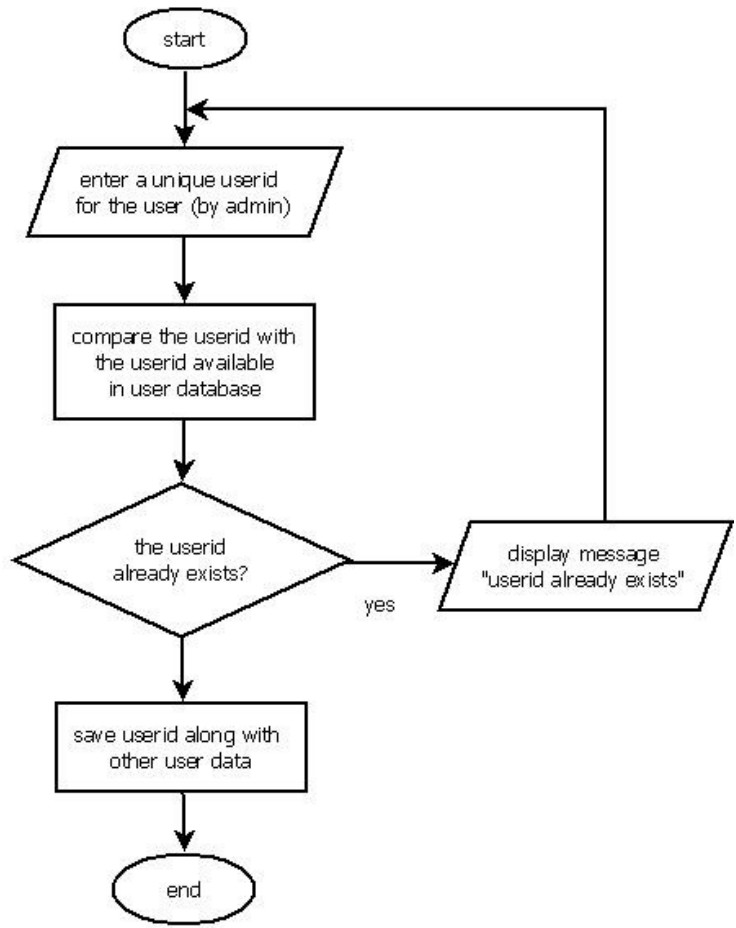

**Note**: As the user-id is created by the MIS administrator, the above flow chart is a part of the admin module.

## 4. Implementing the login process

The type and level of access to an MIS will be typically controlled by the username and password. The login module will check the validity of the username and password and determine the level of access. Based on the access rights of the user (as described above) the login process will allow a unique combination of rights to view/edit specific areas of the MIS database.

### Different access rights for different User Login

The MIS login process verifies the role of a user which may be staff or manager or administrator and presents different types of menu such as the staff menu or manager menu or administrator menu accordingly.

# Difference in access to different components based on user login

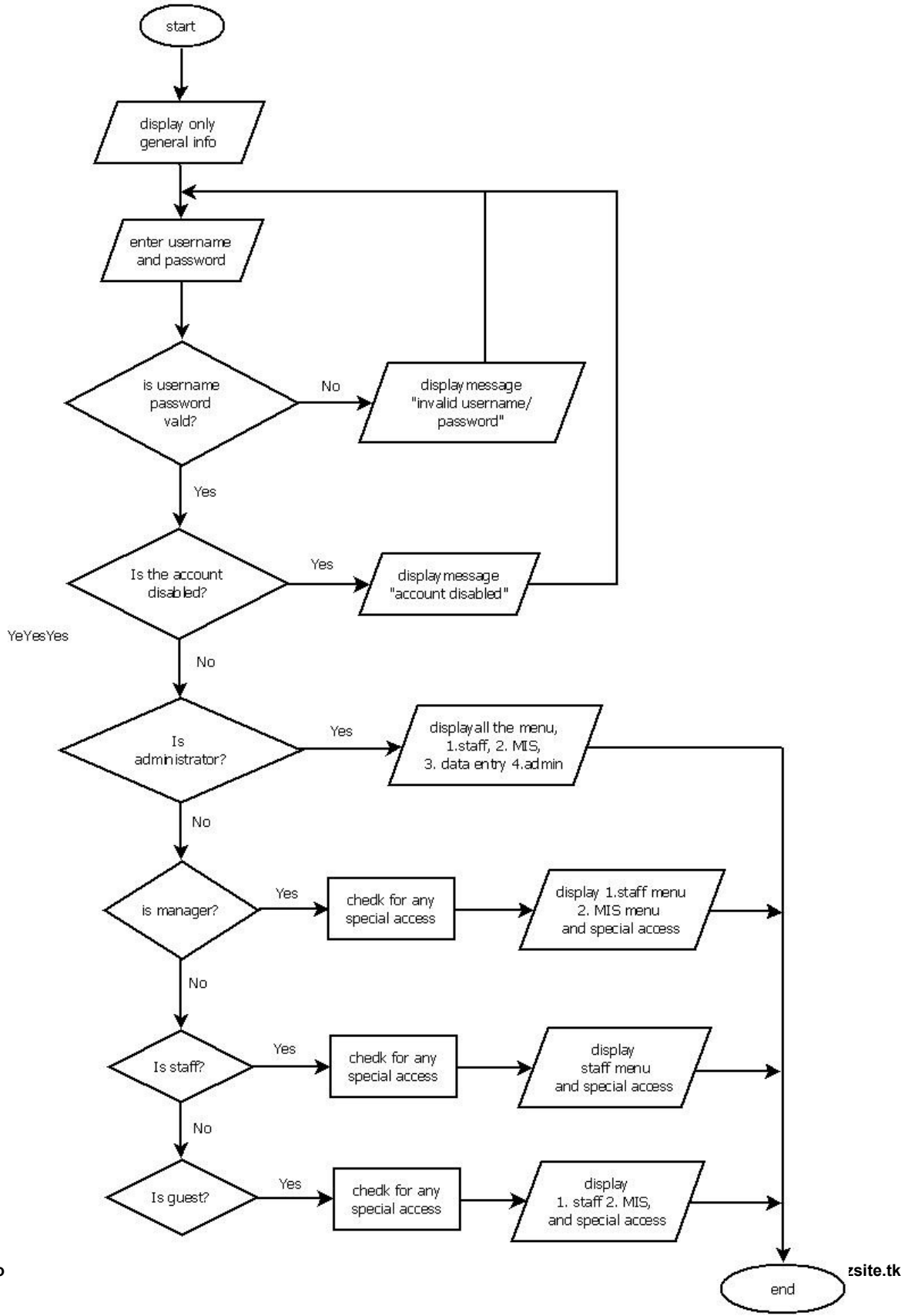

### Changing and recovering password

**The user may have rights to change his password**
The system may ask the user to enter the old password and new password. If the old password matches with the actual old password the system may change the password to the entered new password.

**The user may have rights to recover a forgotten password**
If the user forgets his password, the system may prompt for a hint question and expect a hint answer. If the hint answer provided by the user matches with the hint answer entered by the user, then the user may be issued a new password.

## 5. Summary

The MIS data is critical to an organization and should be protected from misuse by wrong persons. Although The MIS data is typically meant for the senior managers each MIS report may not be required by every manager. The access to MIS data is determined by the role of an individual in the organization and controlled by the MIS administrator accordingly. The access is generally determined by the following parameters, (a) the type of user (such as staff or manager etc.), (b) the type of data (whether general data or managerial data), (c) level of access (read/ write/ admin access) and (d) special access allocated by MIS admin. By combining all the above four parameters, each individual user can be allocated exact specific rights required to access the MIS. The method of allocating "special rights" takes care of all the exceptions required from the rights of a group.

6. Gorry and Morton, *A framework for Management Information systems*, http://archive.org/details/frameworkformana00gorr

7. Laudon and Laudon, *Management Information Systems*,

8. Gorden Davis, *Management Information Systems*,

9. University of Mumbai, *Introduction to Management Information Systems*, http://www.mu.ac.in/mis

10. Craig Larman, *Applying UML and Patterns*, Prentice Hall PTR

11. Umakant Mishra, *Using TRIZ for Anti-Virus Development- Building better software through Continuous Innovation*, 2013, http://www.amazon.com/Using-TRIZ-Anti-Virus-Development-Continuous/dp/9351268829

12. Daniel Amor, *E-business Revolution*, 2000, PTR Prentice Hall,